\documentclass[preprint,showpacs,superscriptaddress, aps]{revtex4}
\usepackage{graphicx}
\begin{document}
\title{Fano resonance in electron transport through parallel double quantum dots
 in the Kondo regime}
\author {Guo-Hui Ding}
\affiliation{Institute of Physics and Applied Physics, Yonsei
University, Seoul 120-749, Korea}
\author{Chul Koo Kim}
\affiliation{Institute of Physics and Applied Physics, Yonsei
University, Seoul 120-749, Korea}
\author{Kyun Nahm}
\affiliation{Department of Physics, Yonsei University, Wonju
220-710, Korea}


\begin{abstract}
 Electron transport through parallel double quantum dot system  with
interdot tunneling  and  strong on-site Coulomb interaction is
studied in the Kondo regime by using the finite-$U$ slave boson
technique.   For a system of quantum dots with degenerate energy
levels, the linear conductance reaches the unitary limit
($2e^2/h$) due to the Kondo effect at low temperature when the
interdot tunneling is absent. As the interdot tunneling amplitude
increases, the conductance decreases in the singly occupied regime
and a conductance plateau structure appears. In the crossover to
the doubly occupied regime, the conductance increases to reach the
maximum value of $G=2e^2/h$. For parallel double dots with
different energy levels, we show that the interference effect
plays an important role in the electron transport. The linear
conductance is shown to have an asymmetric line shape of the Fano
resonance as a function of gate voltage.

\end{abstract}
 \maketitle
\newpage

\section{introduction}
Due to the wave nature of electrons and the confined geometries in
mesoscopic systems,  the interplay between interference and
interaction becomes one of  the central issues in mesoscopic
physics.  Preservation of quantum coherence in electron transport
through an interacting regime has been manifested  in the observed
Kondo effect in semiconductor quantum dot systems\cite{1}, and
more explicitly in the conductance AB oscillation in the
interference experiment with a quantum dot embedded in one arm of
an Aharonov-Bohm(AB) interferometer\cite{2}. Recently, the Fano
resonance has attracted much research interest as another
important interference effect in mesoscopic systems . The Fano
effect was first proposed as a result of the interference between
resonant and non-resonant processes in the field of atomic
physics\cite{3}. It is found to be a ubiquitous phenomenon
observed in a large variety of experiments including neutron
scattering, atomic photoionization, Raman scattering, and optical
absorption. One recent progress is the observation of the Fano
resonances in condensed matter systems, including an impurity atom
on metal surface\cite{4}, single-electron transistor\cite{5,6},
quantum dot in AB interferometer\cite{7,8}.

 In this paper we show the Fano effect, which can be manifested by gate voltage
dependence of the linear conductance, is also important for the
electron transport through double quantum dots(DQDs) in parallel
configuration. For electron tunneling through quantum dots, it is
well known that the strong on-site coulomb interaction leads to
the Kondo effect at low temperatures, so that the coexistence of
the Fano resonance with the Kondo effect is expected to yield
interesting transport phenomena. Electron transport through DQDs
with series\cite{9} and parallel\cite{10} configurations have been
realized in experiments, through which studies on the molecular
states of the double dots and also the interference effect are
carried out. Most of theoretical studies\cite{11,12,13,14} are
devoted to electron transport through DQDs connected in series,
while relatively little attention is paid for the parallel
configuration case especially for the system in the Kondo
regime\cite{13}. For the DQD system with a parallel coupling,
interference effect should play an important role. Thus, in order
to understand the role of the Fano effect, it is essential to take
into account the coherence of the whole system. A model of the
electron transport through a closed AB interferometer containing
two single level quantum dots, which assumes the electrons
transport through quantum dots are in full coherence, has been
investigated in Ref.[15].Interesting phenomena such as
flux-dependent level attraction and interference induced
suppression of conductance have been found. But the effect of
one-site Coulomb interaction and the interdot tunneling haven't
been considered. Ghost Fano resonance has also been observed in
the study of electron transport through  parallel DQDs with
interdot tunneling but no on-site Coulomb interaction\cite{16}.

In this paper we shall investigate electron transport through
parallel DQDs(schematically plotted in Fig.1) with interdot
tunneling and on-site Coulomb interaction using the finite-$U$
slave boson mean field theory(SBMFT) approach developed by Kotliar
and Ruckenstein\cite{17}. This formulation reproduces the results
derived from the well known Gutzwiller variation wave function at
zero temperature, and therefore is believed to be a powerful tool
to study strong correlation effect of electron systems. The
finite-U SBMFT has already been applied to investigate electron
transport through single quantum dot\cite{18}, DQDs in series in
the Kondo regime \cite{19} and persistent current in mesoscopic
ring\cite{20}, and found to give good quantitative results for the
Kondo effect on the linear conductance.

\section{The finite-U slave boson mean field theory of parallel-coupled DQDs}
  Electron transport through parallel DQDs with interdot
tunneling and on-site Coulomb interation can be described by the
following Anderson impurity model:
\begin{equation}
H=\sum_{k\eta\sigma} \epsilon_{k\eta\sigma}
c^\dagger_{k\eta\sigma} c_{k\eta\sigma}+\sum_{i}\epsilon_i
d^\dagger_{i\sigma}d_{i\sigma}+\sum_{i}U
n_{di\uparrow}n_{di\downarrow}+
t_c\sum_\sigma(d^\dagger_{1\sigma}d_{2\sigma}+d^\dagger_{2\sigma}d_{1\sigma})
+\sum_{k\eta\sigma i}
(v_{\eta i}d^\dagger_{i\sigma}c_{k\eta\sigma}+H.c.)\;,\\
\end{equation}
where $c_{k\eta\sigma}(c^\dagger_{k\eta\sigma})$ denote
annihilation(creation) operators for electrons in the
leads($\eta=L, R$),  and $d_{i\sigma}(d^\dagger_{i\sigma})$ those
of the single level state in the i-th dot( $i=1,2$). U is the
intra-dot coulomb interaction between electrons, $t_c$ is the
interdot tunnel coupling, and  $v_{\eta i}$ is the tunnel matrix
element between lead $\eta$ and dot $i$.   We consider the
symmetric coupling case with  $\Gamma_i^L=\Gamma_i^R=\Gamma_i$ ,
where $\Gamma_i^\eta=2\pi\sum_k |v_{\eta i}|^2
\delta(\omega-\epsilon_{k\eta\sigma})$ is the hybridization
strength between the $i$-th dot and the lead $\eta$.

    In the finite-$U$ slave boson approach\cite{17, 18}, a set of auxiliary bosons $e_i, p_{i\sigma}, d_i$ are introduced
for each dot,  which act as projection operators onto the empty,
singly occupied(with spin up and spin down), and doubly occupied
electron states on the quantum dot, respectively.  The fermion
operators $d_{i\sigma}$ are replaced by $d_{i\sigma}\rightarrow
f_{i\sigma}z_{i\sigma} $, with
$z_{i\sigma}=e^\dagger_ip_{i\sigma}+p^\dagger_{i\bar\sigma}d_i$.
In order to eliminate unphysical states, the following constraint
conditions are imposed :$\sum_{\sigma}
p^\dagger_{i\sigma}p_{i\sigma}+e^\dagger_i e_i+d^\dagger_id_i=1$,
and
$f^\dagger_{i\sigma}f_{i\sigma}=p^\dagger_{i\sigma}p_{i\sigma}+d^\dagger_id_i(\sigma=\uparrow,
\downarrow)$. Therefore,  the Hamiltonian (1) can be rewritten as
the following effective Hamiltonian in terms of the auxiliary
boson $e_i, p_{i\sigma}, d_i$  and the pesudo-fermion operators
$f_{i\sigma}$:
\begin{eqnarray}
H_{eff}&=&\sum_{k\eta\sigma} \epsilon_{k\eta\sigma}
c^\dagger_{k\eta\sigma} c_{k\eta\sigma}+\sum_{i=1,2}\epsilon_i
f^\dagger_{i\sigma}f_{i\sigma}+\sum_{i} Ud^\dagger_id_i+
t_c\sum_\sigma(
z^\dagger_{1\sigma}f^\dagger_{1\sigma}f_{2\sigma}z_{2\sigma}+H.c)
\nonumber\\
& +&\sum_{k\eta\sigma i} (v_{\eta i}z^\dagger_{i\sigma}
f^\dagger_{i\sigma}c_{k\eta\sigma}+H.c.) +\sum_{i}
\lambda^{(1)}_i(\sum_{\sigma}
p^\dagger_{i\sigma}p_{i\sigma}+e^\dagger_i e_i+d^\dagger_id_i-1)
\nonumber\\
&+&\sum_{i\sigma}\lambda^{(2)}_{i\sigma}(f^\dagger_{i\sigma}f_{i\sigma}-p^\dagger_{i\sigma}p_{i\sigma}-d^\dagger_id_i
)\;,
\end{eqnarray}
 where the constraints are incorporated by the Lagrange multipliers $\lambda^{(1)}_i$ and
$\lambda^{(2)}_{i\sigma}$. The first constraint can be interpreted
as a completeness relation of the Hilbert space in each dot, and
the second one equates the two ways of counting the fermion
occupancy of a given spin\cite{17}. In the framework of the
finite-U SBMFT, the slave boson operators $e_i, p_{i\sigma}, d_i $
and the parameter $z_\sigma$ are replaced by real c numbers. In
this paper, we only consider the spin degenerate case without
external magnetic field, so that all parameters are independent of
the electron spin. We can neglect the spin index $\sigma$ in the
parameters hereafter. Thus in the mean field approximation, the
effective Hamiltonian is given as
\begin{eqnarray}
H^{MF}_{eff}=&=&\sum_{k\eta\sigma} \epsilon_{k\eta}
c^\dagger_{k\eta\sigma}
c_{k\eta\sigma}+\sum_{i=1,2}{\tilde\epsilon_i}f^\dagger_{i\sigma}f_{i\sigma}
+{\tilde t_c}\sum_\sigma(f^\dagger_{1\sigma}f_{2\sigma}+H.c)
\nonumber\\
& +&\sum_{k\eta\sigma i} ({\tilde v}_{\eta i}
f^\dagger_{i\sigma}c_{k\eta\sigma}+H.c.)+E_g\;,
\end{eqnarray}
where ${\tilde t_c}=t_cz_1z_2$ and ${\tilde v_{\eta i}}=v_{\eta
i}z_{i}$ represents the renormalized tunnel coupling between
quantum dots and the renormalized tunnel amplitude between $i$-th
quantum dot and the lead $\eta$, respectively. $z_1$ and $z_2$ can
be regarded as the wave function renormalization factors in the
quantum dots. ${\tilde\epsilon_i}=\epsilon_i+\lambda^{(2)}_i$ is
the renormalized dot energy level and $E_g=\sum_{i}
[\lambda^{(1)}_i(
2p^2_{i}+e^2_i+d^2_i-1)-2\lambda^{(2)}_i(p^2_{i}+d^2_i)+Ud^2_i]$
is an energy constant.

  Within this mean field effective Hamiltonian (3) the current formula through the DQDs is given
as\cite{15}
\begin{equation}
I={e\over h}\sum_{\sigma}\int d\omega
[n_L(\omega)-n_R(\omega)]T(\omega)\;,\\
\end{equation}
where the transmission probability $T(\omega)=Tr[G^a(\omega)
{\tilde\Gamma}^R G^r(\omega){\tilde\Gamma}^L]$, and
${\tilde\Gamma^L}={\tilde\Gamma}^R=
\left ( \begin{array}{cc} {\tilde\Gamma}_1 & \sqrt{{\tilde\Gamma}_1{\tilde\Gamma}_2  } \\
\sqrt{{\tilde\Gamma}_1{\tilde\Gamma}_2  } & {\tilde\Gamma}_2
\end{array} \right )$, with ${\tilde\Gamma}_i=z^2_i\Gamma_i$. The
retarded/advanced  Green's functions(GF) $G^{r/a}(\omega)$ have
$2\times 2$ matrix structures, which account for the double dot
structure of the system. The matrix elements of the retarded GF
are defined in time space as
$G_{ij}^r(t-t')=-i\theta(t-t')<\{f_{i\sigma}(t),
f^+_{j\sigma}(t')\}>$.  By applying the equation of motion
method\cite{21}, one can obtain the retarded GF explicitly as
\begin{equation}
G^r(\omega)=\left ( \begin{array}{cc}
 \omega-{\tilde\epsilon_1}+i{\tilde\Gamma}_1 & -{\tilde t_c}+i\sqrt{{\tilde\Gamma}_1{\tilde\Gamma}_2  } \\
-{\tilde t_c}+i\sqrt{{\tilde\Gamma}_1{\tilde\Gamma}_2  } & \omega-{\tilde\epsilon_2}+i{\tilde\Gamma}_2
\end{array}\right )^{-1}\;,\\
\end{equation}
The advanced GF is given by $G^a(\omega)=[G^{r}(\omega)]^\dagger$.
Substituting the retarded/advanced GF to the formula of
transmission probability, one obtains
\begin{equation}
T(\omega)={[{\tilde \Gamma}_1(\omega-{\tilde \epsilon}_2)+{\tilde
\Gamma}_2(\omega-{\tilde \epsilon}_1)+2{\tilde
t_c}\sqrt{{\tilde\Gamma}_1{\tilde\Gamma}_2  }]^2\over {
[(\omega-{\tilde \epsilon}_1)(\omega-{\tilde \epsilon}_2)-{\tilde
t_c}^2]^2+[{\tilde \Gamma}_1(\omega-{\tilde \epsilon}_2)+{\tilde
\Gamma}_2(\omega-{\tilde \epsilon}_1)+2{\tilde
t_c}\sqrt{{\tilde\Gamma}_1{\tilde\Gamma}_2  }]^2 }}\;.\\
\end{equation}
The  conductance $G$ at the absolute zero temperature in the limit
of zero bias voltage is given by
$G={dI\over{dV}}|_{V=0}={2e^2\over h}T(\omega=0)$.

It is noticed that the formula for the transmission probability
and conductance is equivalent to that of the transport through
non-interacting double QD system, except that, in this case, the
dot levels ${\tilde\epsilon}_i$, the coupling strength
${\tilde\Gamma}_i$ and ${\tilde t_c}$ are renormalized. Therefore,
the electron transport through DQDs is characterized by the
parameters ${\tilde\epsilon}_i$, ${\tilde\Gamma}_i$ and ${\tilde
t_c}$.  However, it should be noted that ${\tilde\epsilon}_i$,
${\tilde\Gamma}_i$ and ${\tilde t_c}$ show strong dependence on
the gate voltage applied to the quantum dots, hence the result of
linear conductance is quite different from the non-interacting
model.  In the spin degenerate case, we have ten unknown
parameters $e_i, p_i, d_i, \lambda^{(1)}_i,
\lambda^{(2)}_i$(i=1,2) in total to determine. From the
constraints and the equation of motion of the slave boson
operators in the effective Hamiltonian, we obtain one set of
self-consistent equations, which is a straightforward generalized
form of the single dot case as discussed in Ref.[18].
 In this set of equations, the distribution GF of the quantum dots $G_{ij}^<(t-t')=i<
f^+_{j\sigma}(t')f_{i\sigma}(t)>$ is involved, and its Fourier
transform is given by
$G^<(\omega)=iG^r(\omega)[{\tilde\Gamma^L}n_L(\omega)+{\tilde\Gamma^R}n_R(\omega)]G^a(\omega)$.
We have solved the self-consistent equations numerically.

 In the following, we discuss the result of our calculation.
First, we consider two identical QDs case:
$\epsilon_1=\epsilon_2=\epsilon_d$ and $\Gamma_1=\Gamma_2$.
Following Eq.(5), the transmission probability in this case has a
Breit-Wigner resonance form,
\begin{equation}
T(\omega)={4{\tilde \Gamma}^2\over{(\omega-{\tilde
\epsilon_d}-{\tilde t_c})^2 +4{\tilde \Gamma}^2} }\;.\\
\end{equation}
The retarded GF on each dot is also explicitly given as
\begin{equation}
G_{ii}^r(\omega)={1\over 2}[{1\over {\omega-({\tilde
\epsilon_d}-{\tilde t_c})+0^+}}+{1\over {\omega-({\tilde
\epsilon_d}+{\tilde t_c})+2i{\tilde\Gamma}}}        ]\;.\\
\end{equation}
The spectral density in the $i$-th QD follows from the relation
$\rho_i(\omega)=-Im G^r_{ii}(\omega+i0^+)/\pi$. It shows that the
spectral density is the sum of one Lorentizan with the peak
position located at
${\tilde\epsilon}_{bond}=\tilde\epsilon_d+{\tilde t_c}$ and one
Dirac $\delta$ peak at
${\tilde\epsilon}_{antibond}=\tilde\epsilon_d-{\tilde t_c}$, where
${\tilde\epsilon}_{bond}$ and ${\tilde\epsilon}_{antibond}$
corresponds to energy of the bonding and the antibonding state of
quantum dots, respectively. The bonding state of DQDs has level
broadening  $2{\tilde\Gamma}$ due to its coupling with the leads.
The $\delta$ peak structure indicates that the antibonding state
is totally decoupled from the leads. Therefore the electrons
transport only through the channel of the bonding state, which
gives a Breit-Wigner resonance form in the transmission.

  In Fig.2 we study the effect of interdot tunneling on the transmission
probability and the local density of state of QD in the singly
occupied regime. Here we take the hybridization strength as the
energy unit $\Gamma=1$, and $\epsilon_d=-2$.  Fig.2(a) shows that
with increasing the interdot  coupling $t_c$, the line shape of
Breit-Wigner resonance of transmission is preserved, while the
center of the resonance shifts to higher energy. Thus, the value
of transmission probability at zero frequency $T(\omega=0)$
decreases, which, in turn,  results in suppression of the linear
conductance at zero bias voltage. For local density of state shown
in Fig.2(b),  we see that the antibonding state energy is always
nearby the Fermi energy of the lead, whereas the center of
spectral density contributed from the bonding state shifts to
higher energy along with increasing $t_c$.

 The linear conductance $G$ as a function of
the energy level $\epsilon_d$ of the QD at zero temperature is
plotted in Fig.3(a) for several values of interdot tunneling
$t_c$. When there is no direct tunneling between two dots
($t_c=0$), the conductance reaches the unitary limit($G=2e^2/h$)
in the Kondo regime as expected. With increasing tunnel coupling
$t_c$, the conductance becomes suppressed and forms a plateau
structure in the regime of  singly occupied QD state. When QDs
cross over to the doubly occupied state regime, the conductance
increases to the maximum value $G=2e^2/h$. The line shape of the
linear conductance can be explained from the gate voltage
dependence of  the spectral density and the zero frequency
transmission of QD. In Fig.3(b) we plot the bonding state energy
${\tilde\epsilon}_{bond}$, the antibonding state energy
${\tilde\epsilon}_{antibond}$ and the level broadening
$2{\tilde\Gamma}$ as functions of $\epsilon_d$ with $t_c=1$. One
can see that in the singly occupied regime with decreasing
$\epsilon_d$, the antibonding state energy
${\tilde\epsilon}_{antibond}$ is fixed around the Fermi energy of
the leads($\epsilon_F=0$). It indicates that
${\tilde\epsilon_d}\approx{\tilde t_c}$, and the conductance
$G/(2{e^2/ h})={4{\tilde \Gamma}^2/[({\tilde \epsilon_d}+{\tilde
t_c})^2 +4{\tilde \Gamma}^2] }\approx {1/ ({\tilde
t_c}^2/{\tilde\Gamma}^2+1)}$. For this identical quantum dot case,
the value of ${\tilde t_c}^2/{\tilde\Gamma}^2$ is given by its
bare value ${\tilde t_c}^2/{\tilde\Gamma}^2={ t_c}^2/{\Gamma}^2$.
Consequently, the conductance shows a plateau structure and the
ratio of $t_c/\Gamma$ determines the height of the conductance
plateau. This is in agreement with the value of conductance at the
plateau structure for different $t_c$ as shown in Fig.3(a). With
$\epsilon_d$ deceasing further, the QD state crosses over from the
singly occupied to the doubly occupied regime, and
${\tilde\epsilon}_{bond}$  goes through from positive value to
negative value. At the point,
${\tilde\epsilon}_{bond}=\tilde\epsilon_d+{\tilde t_c}=0$, we
obtain the maximum conductance $G=2e^2/h$. Further down, the level
broadening $2\tilde \Gamma$ approaches zero and the DQD will be
totally decoupled from the leads, thus the conductance becomes
zero.

Next, we consider DQD system with different dot levels,
$\epsilon_1\neq\epsilon_2$, and define
$\bar\epsilon=(\epsilon_1+\epsilon_2)/2$ and
$\Delta\epsilon=\epsilon_1-\epsilon_2$.  For the sake of
simplicity, we still assume $\Gamma_1=\Gamma_2=\Gamma$. It is
noted that in this case the renormalized hybridization strength
${\tilde\Gamma}_1\neq{\tilde\Gamma}_2$. In Fig.4 we plot the
linear conductance as a function of average energy of the dot
levels. The parameters $\Delta \epsilon=0.5, 1.0 $ and  $t_c=0.0,
0.5, 1.0, 1,5$ are used.   In the case of DQDs without direct
tunnel coupling $t_c=0$, Fig.4(a) shows that the conductance curve
has a narrow dip around the point $\bar\epsilon=-U/2$. From the
formula for transmission Eq.(6), we note that the conductance
vanishes when the condition ${\tilde \Gamma}_1{\tilde
\epsilon}_2+{\tilde \Gamma}_2{\tilde \epsilon}_1=2{\tilde
t_c}\sqrt{{\tilde\Gamma}_1{\tilde\Gamma}_2 } $ is satisfied.  The
strictly zero transmission is a consequence of destructive quantum
interference for electron transport through the parallel DQDs, and
it is absent for systems with DQDs connected in series. It is
interesting to notice that only when the DQDs with different
energy levels $\Delta\epsilon\neq 0$, this characteristic of
interference is revealed. It originates from the fact that, in
this case, both the bonding and antibonding state channels are
involved in the transmission. As the energy difference
$\Delta\epsilon$ increases, the dip becomes more broadened. For
non-zero interdot tunnel couplings as shown in Fig.4(b), (c) and
(d), the conductance curves have asymmetric line shapes, which are
typical for the Fano resonance.   This results from the
constructive and destructive interference processes for electrons
transmitted through the channels of bonding and antibonding
states. It is noted that line broadening of the Fano dip or peak
depends on the value of dot level difference $\Delta\epsilon$,
which is similar to the non-interacting DQDs case\cite{16}. The
effect of on-site interaction $U$ is to introduce strong
renormalization of the dot levels and the hybridization strength,
hence the center of the Fano resonance and line broadening have
nonlinear dependence on the interdot tunneling $t_c$ and the level
difference $\Delta\epsilon$. It is interesting to notice the Fano
resonances obtained in this study have some similarity with the
experiment results in Ref.\cite{5}. Although their experiment is
on electron transport through single QD, the coupling strength
between quantum dot and the lead is strong and the Kondo effect
and multilevel of QD might be involved in the electron transport.
Recently, B${\ddot u}$sser et al.\cite{22} have studied the
electron transport through multilevel quantum dots using
exact-diagonalization techniques. It is interesting to notice that
they have also found conductance dip structure induced by
interference effect as shown in Fig.4(a). Actually,  when $t_c=0$,
the model studied in our paper is equivalent to considering two
levels in single quantum dot.

For a DQD system with energy level difference, the local density
of state in $i-$th QD$(i=1,2)$ is given by
\begin{equation}
\rho_i(\omega)={[\sqrt{{\tilde \Gamma}_i}(\omega-{\tilde
\epsilon}_{\bar i})+\sqrt{{\tilde \Gamma}_{\bar i}}{\tilde
t_c}]^2\over { [(\omega-{\tilde \epsilon}_1)(\omega-{\tilde
\epsilon}_2)-{\tilde t_c}^2]^2+[{\tilde \Gamma}_1(\omega-{\tilde
\epsilon}_2)+{\tilde \Gamma}_2(\omega-{\tilde
\epsilon}_1)+2{\tilde
t_c}\sqrt{{\tilde\Gamma}_1{\tilde\Gamma}_2  }]^2 }}\;.\\
\end{equation}
In Fig.5. we plot the local density of state in each dot.  The
line shape of density of state can be regarded as a superposition
of a Fano line shape close to the antibonding state energy and a
Breit-Wigner resonance around the bonding state energy(see
Ref.[16] for detail discussion). The interference effect on the
local density of state is manifested clearly as compared with that
in Fig.2(b).
\section{summary}
 In summary, we have studied the electron transport through DQDs in parallel
configuration with interdot tunneling in the Kondo regime. The
strong Coulomb repulsion in the dots is taken into account via the
finite-$U$ slave boson technique. The results of our calculation
indicate several distinct features from the non-interacting
model\cite{16}: The conductance shows plateau structure as a
function of the dot level in the singly occupied regime; Without
interdot tunneling $t_c=0$, there is dip structure on the
conductance plateau when the energy levels of two dots are
different. When $t_c\neq 0$, the conductance has Fano resonance
line shape on the conductance plateau as a function of the
averaged dot level; The energies of the bonding and antibonding
states and the level broadening of the bonding state are strongly
renormalized compared to the noninteracting model case. For
instance, the antibonding state energy is almost fixed around the
Fermi energy of the lead in the singly occupied region.  The
results are also different from that of the DQDs in series, in
which the maximum conductance is achieved when the interdot
tunneling $t_c=1.0$ and no Fano resonance is
observed\cite{11,12,13,14, 19}. The Fano effect for parallel DQDs
originates from the interference effect for electron transport
through the two channels of bonding and antibonding states of
parallel DQDs.  In one recent experiment, Chen et al.\cite{10}
have studied the Kondo effect in parallel DQDs system. But the
maximum conductance obtained in their experiment is only about
$0.1 e^2/h$ by varying the gate voltage and interdot tunneling, so
that we think the full coherent electron transport through DQDs
isn't achieved and the interference effect is not manifested. One
may expect further experiments on the parallel DQDs system can
observe the conductance plateau structure and also the Fano
resonance as discussed above.
\begin{acknowledgments}
This research was supported by BK21 project and Korea Research
Foundation(KRF-2003-005-C00011).
\end{acknowledgments}

\newpage
\begin{figure}[htp]
\includegraphics[width=0.8\columnwidth ]{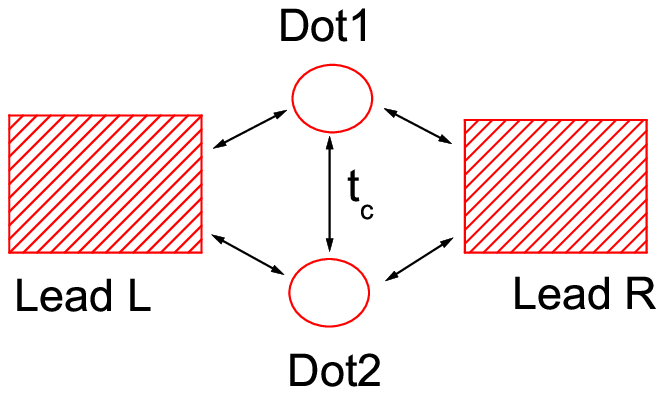}
\caption{ Parallel double quantum dots with interdot tunneling
$t_c$.}
\end{figure}
\begin{figure}[htp]
\includegraphics[width=0.8\columnwidth ]{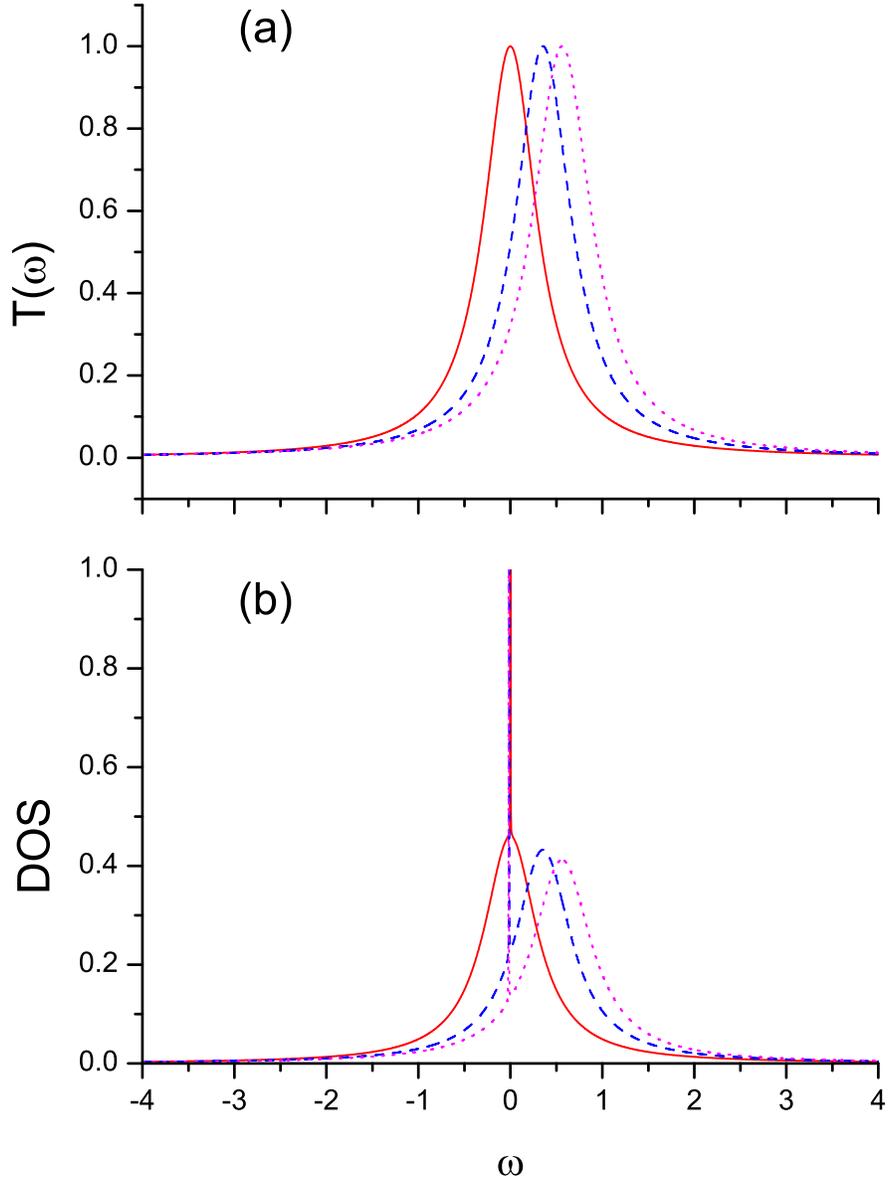}
\caption{(a) The transmission probability $T(\omega)$ and (b) the
local density of state for the system with two identical quantum
dots. Used parameters are $U=4.0, \Gamma^L=\Gamma^R=1.0,
\epsilon_d=-2.0$. The interdot tunnel coupling $t_c$ are
$0.0$(solid line), $0.5$(dashed line), $1.0$(dotted line). (We
take the energy unit as $\Gamma=1$, and $\omega=0$ corresponds to
the Fermi energy of the leads).}
\end{figure}
\begin{figure}[htp]
\includegraphics[width=0.8\columnwidth ]{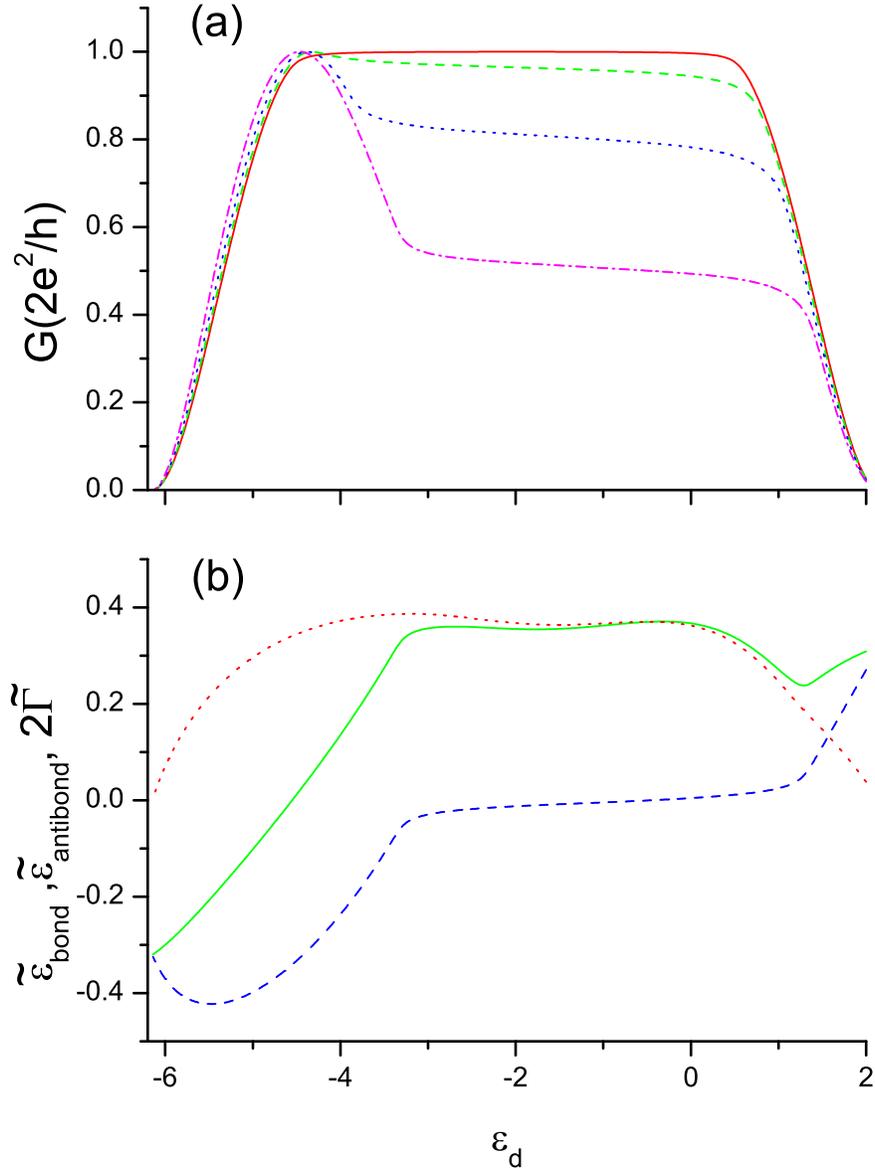}
\caption{(a)The linear conductance as a function of the dot level
at zero temperature. Used parameters are $U=4.0,
\Gamma^L=\Gamma^R=1.0$, and the interdot tunneling $t_c=0.0$(solid
line), $0.2$(dashed line), $0.5$(dotted line), $1.0$(dash-dotted
line). (b) The bonding state energy $\tilde\epsilon_{bond}$(solid
line), the antibonding state energy
$\tilde\epsilon_{antibond}$(dashed line) and the level broadening
$ 2\tilde\Gamma$ of the bonding state(dotted line)  for
$t_c=1.0$.}
\end{figure}
\begin{figure}[htp]
\includegraphics[width=0.8\columnwidth ]{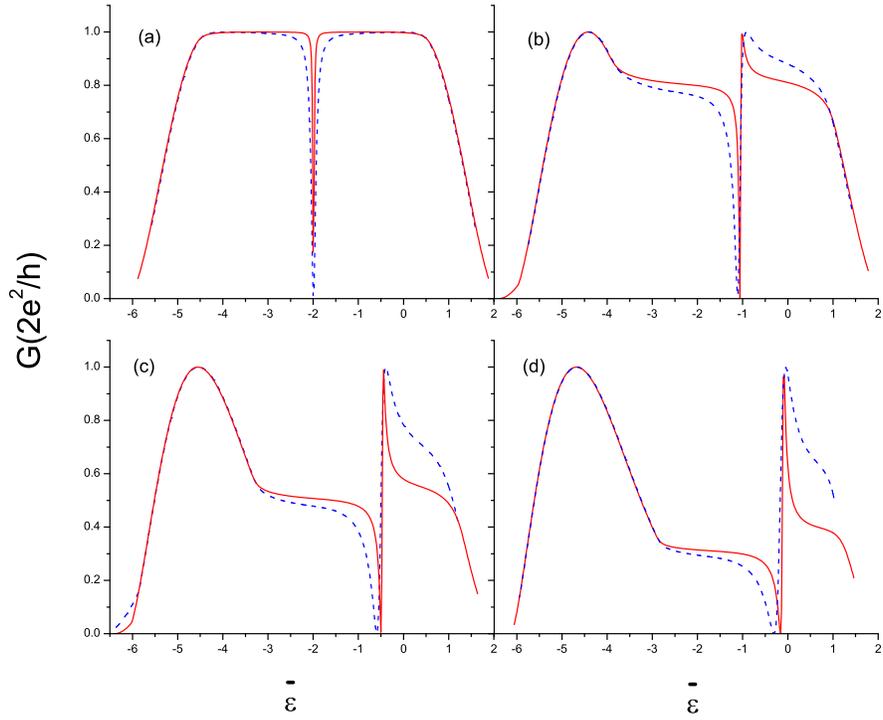}
\caption{ The linear conductance at zero temperature for parallel
double quantum dots with different energy levels. Used parameters
are $U=4.0, \Gamma^L=\Gamma^R=1.0$. The energy level differences
are $\Delta\epsilon=0.5$ (solid line), $1.0$ (dashed line).  (a),
(b), (c), and (d) correspond to the interdot tunneling $t_c=0.0,
0.5, 1.0, 1.5 $, respectively.}
\end{figure}
\begin{figure}[htp]
\includegraphics[width=0.8\columnwidth]{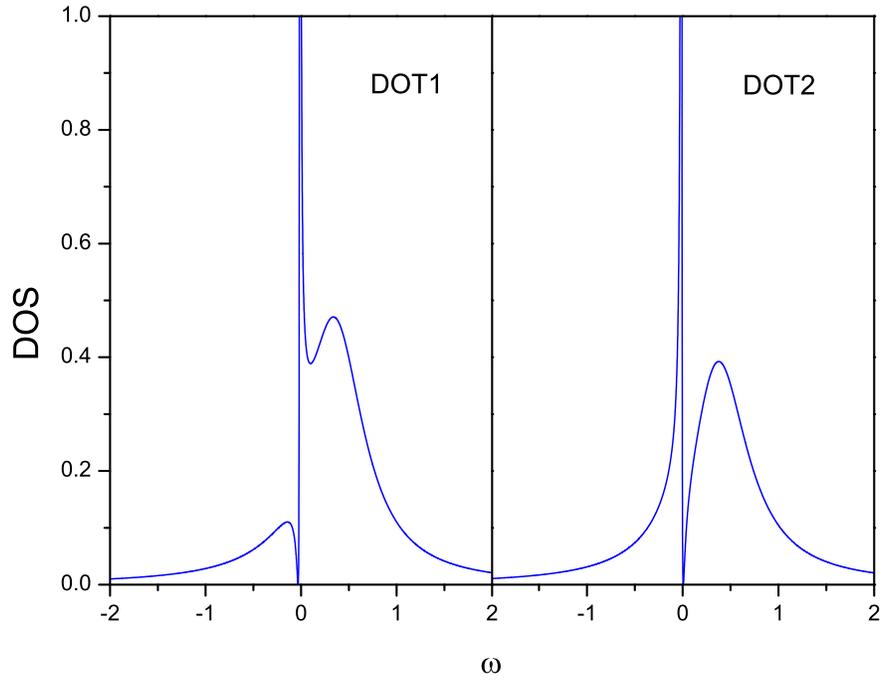}
\caption{ The local density of state for each quantum dot. Used
parameters are $U=4.0, {\bar\epsilon}=-2.0,\Delta \epsilon=1.0,
t_c=1.0$.}
\end{figure}

\end{document}